\documentclass{amia}
\usepackage[table,xcdraw]{xcolor}
\usepackage{lipsum} %Remove if not needed
\usepackage{rotating}
\usepackage{multirow}
\usepackage{color}
\usepackage{wrapfig} 
\usepackage{threeparttable}
\usepackage{graphicx}
\usepackage{subcaption}
\usepackage{fancyhdr}
\usepackage{amsmath}
\usepackage{amssymb} 
\usepackage{float}
\usepackage{url}
\usepackage{booktabs} 
\usepackage{pifont}             % for dingbats (\ding{51}, \ding{55})
\usepackage[utf8]{inputenc}
\usepackage{amsmath}
\usepackage{hyperref}  % For clickable URLs
\usepackage{tikz}
\usetikzlibrary{calc}
\usetikzlibrary{shapes, arrows, positioning, trees}
\usepackage{pdfpages}
\usepackage{array}    

\begin{document}

\pagenumbering{arabic}

\title{When Documents Disagree: Measuring Institutional Variation in Transplant Guidance with Retrieval-Augmented Language Models}

\author{Yubo Li, MS$^1$, Ramayya Krishnan, PhD$^1$, Rema Padman, PhD$^1$}

\institutes{
    $^1$ Carnegie Mellon University, Pittsburgh, PA, USA
}

\maketitle

\section*{Abstract}

\textit{Patient education materials for solid-organ transplantation vary substantially across U.S. centers, yet no systematic method exists to quantify this heterogeneity at scale. We introduce a framework that grounds the same patient questions in different centers’ handbooks using retrieval-augmented language models and compares the resulting answers using a five-label consistency taxonomy. Applied to 102 handbooks from 23 centers and 1,115 benchmark questions, the framework quantifies heterogeneity across four dimensions: question, topic, organ, and center. We find that 20.8\% of non-absent pairwise comparisons exhibit clinically meaningful divergence, concentrated in condition monitoring and lifestyle topics. Coverage gaps are even more prominent: 96.2\% of question-handbook pairs miss relevant content, with reproductive health at 95.1\% absence. Center-level divergence profiles are stable and interpretable, where heterogeneity reflects systematic institutional differences, likely due to patient diversity. These findings expose an information gap in transplant patient education materials, with document-grounded medical question answering highlighting opportunities for content improvement.}

\section{Introduction}
Large language models (LLMs) are increasingly used to answer medical questions and support patient education, often through retrieval-augmented generation (RAG), in which responses are grounded in external documents such as clinical guidelines or institutional patient education materials. While prior work has focused on improving retrieval quality and reducing hallucinations, less attention has been paid to an upstream issue: can the underlying source documents themselves contain differing guidance? If so, what are the similarities and differences between handbooks on specific topics and what is their provenance?

This question is especially relevant in solid-organ transplantation, where substantial center-level variation is well documented. Analyses of national registry data show that the probability of receiving a deceased-donor kidney transplant within three years of waitlisting varies dramatically across centers within the same donation service area~\cite{king2020major}. These institutional differences extend to patient-facing information: transplant center websites provide incomplete and inconsistent recipient selection criteria~\cite{rivera2025examining}, patient education materials vary widely in readability and quality~\cite{rodrigue2017readability,poudel2024readability}, and a comparative analysis of transplant handbooks using NLP and generative methods found significant variation in the availability and interpretation of clinical guidance across centers~\cite{mace2025improving}.

These findings raise a critical question: when patients ask the same or a similar question to different centers' patient handbook documents, does the resulting advice differ in clinically meaningful ways? If so, content selection for patient handbooks has implications for both clinical and management decision making. In this work, we address this problem by grounding the same patient questions in handbooks from different U.S.\ transplant centers and systematically comparing the resulting responses. Our goal is not to identify a single correct answer but to measure the extent and structure of institutional variation across organs, topics, and centers, contributing to broader efforts to understand the information gaps in transplant patient education within the U.S.\ transplantation system.

\section{Methods}
\subsection{Data Sources and Processing}

Our study draws on two primary data sources: a corpus of transplant patient education handbooks that were generously shared by U.S.\ transplant centers and assembled by the non-profit Transplants.org for analysis, and a curated benchmark of patient questions extracted from multiple patient forums, spanning five solid-organ types. This section describes the collection, scope, and characteristics of each.

\begin{wrapfigure}{r}{0.45\textwidth}
    \centering
    \includegraphics[width=0.5\textwidth, trim={7cm 5cm 7cm 7cm}, clip]{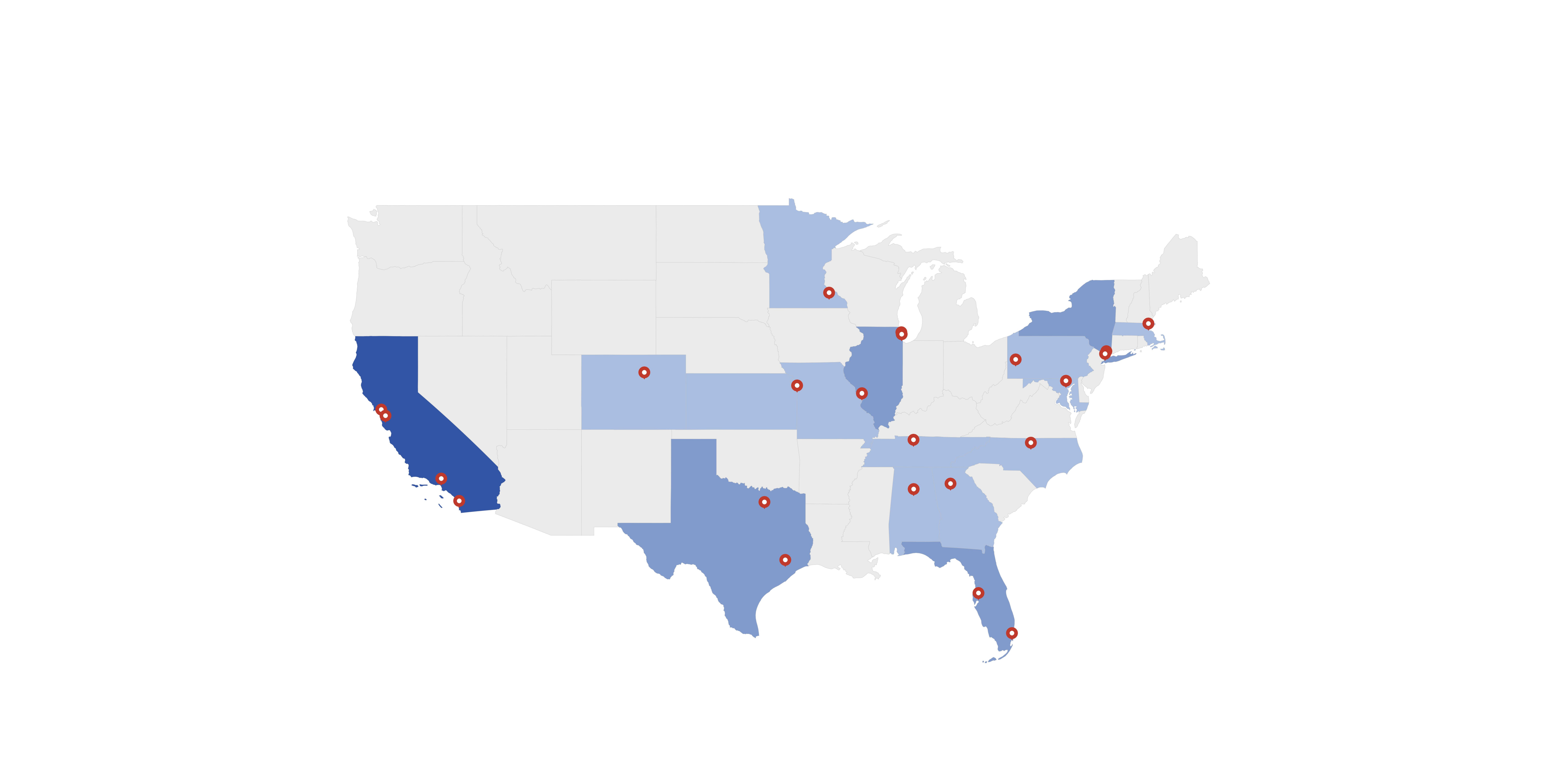}
    \caption{Center distribution map for the handbook corpus. The 23 contributing U.S. transplant centers are geographically dispersed across the country.}
    \label{fig:center_distribution}
\end{wrapfigure}

\textit{Transplant Patient Handbooks.} 
We obtained a corpus of 102 patient education handbooks from 23 major U.S. solid-organ transplant centers, representing 16 of the nation's 20 largest programs by volume. The corpus spans five organ types — heart (26), lung (26), kidney (22), liver (17), and pancreas (11) — and the contributing centers are geographically distributed across the United States (Figure~\ref{fig:center_distribution}), covering both large academic medical centers and community-based transplant programs. All documents were obtained as PDFs from the institutions. Table~\ref{tab:handbook_summary} summarizes the corpus by organ type.

Centers vary in how they organize patient education materials: some provide separate documents for the pre-transplant phase (evaluation, listing, and waiting) and the post-transplant phase (recovery, medications, and long-term follow-up), while others issue a single combined handbook. We treat each phase-specific document as a distinct unit, yielding 37 pre-transplant, 39 post-transplant, and 26 combined handbooks. Each is assigned a unique identifier encoding organ type, institution, and care phase.

\textit{Transplant Patient Question Set.} 
We curated a benchmark of 1{,}115 patient questions to serve as the evaluation set for cross-center comparison. Questions were collected from diverse sources reflecting the real information needs of transplant patients and caregivers. Source types include healthcare institution Q\&A pages (31.2\%), community forums such as Reddit and Mayo Clinic Connect (25.1\%), medical organizations including the National Kidney Foundation and the American Liver Foundation (24.9\%), and other sources including government health agencies and patient advocacy sites (18.8\%).

\begin{table}[H]
\centering
\caption{Summary of the transplant handbook corpus by organ type. \textit{Centers} indicates the number of distinct institutions contributing handbooks for each organ.}
\label{tab:handbook_summary}
\begin{tabular}{lrrrrr}
\toprule
 & \textbf{Heart} & \textbf{Kidney} & \textbf{Liver} & \textbf{Lung} & \textbf{Pancreas} \\
\midrule
Handbooks       & 26 & 22 & 17 & 26 & 11 \\
Centers         & 17 & 14 & 11 & 15 &  8 \\
Pre-transplant  & 10 &  8 &  5 & 10 &  4 \\
Post-transplant & 11 & 10 &  4 & 11 &  3 \\
Combined        &  5 &  4 &  8 &  5 &  4 \\
\bottomrule
\end{tabular}
\end{table}

\begin{figure}[H] 
    \centering
    \begin{subfigure}{0.43\textwidth}
        \includegraphics[width=\textwidth]{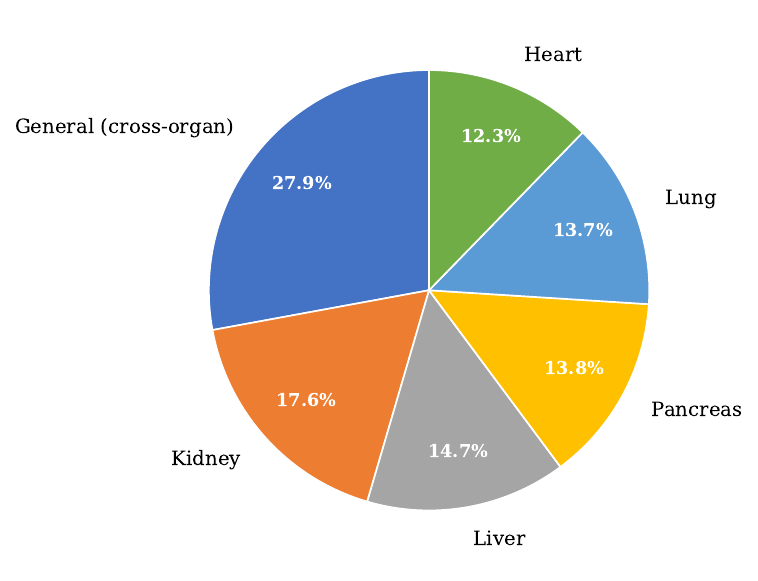}
        \caption{By organ type.}
        \label{fig:question_pie_organ}
    \end{subfigure}
    \hfill
    \begin{subfigure}{0.52\textwidth}
        \includegraphics[width=\textwidth]{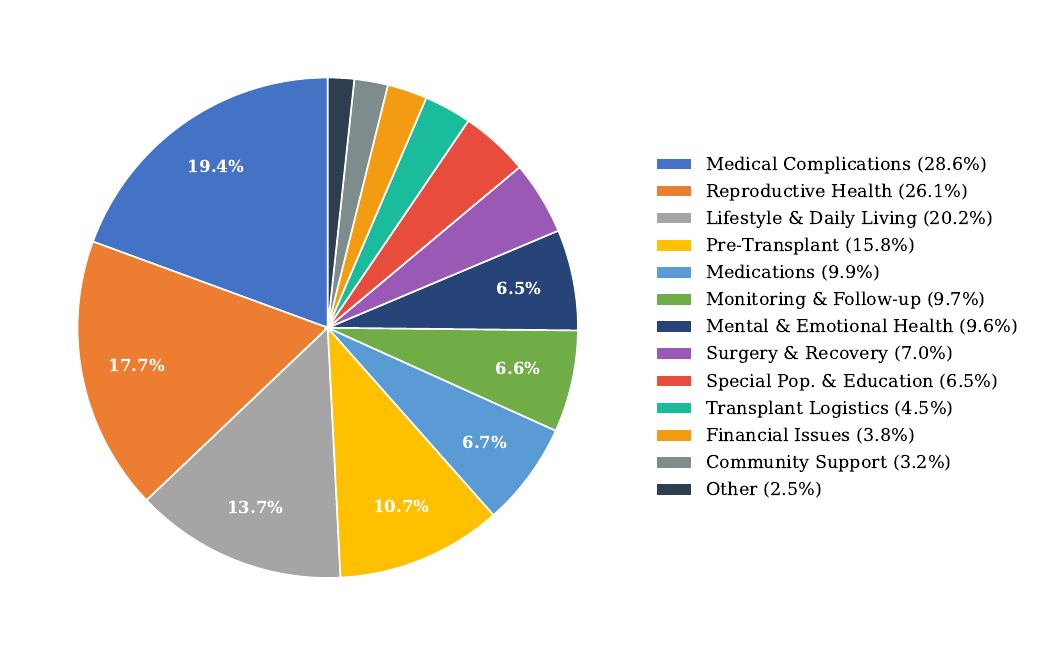}
        \caption{By topic category.}
        \label{fig:question_pie_topic}
    \end{subfigure}
    \caption{Distribution of benchmark questions by organ type and topic category.}
    \label{fig:question_pie}
\end{figure}

Each question is annotated with: (i) an organ type label (heart, kidney, liver, lung, pancreas, or general), (ii) one or more clinical topic categories drawn from a 13-topic taxonomy, and (iii) fine-grained sub-topic tags (43 unique sub-topics). As shown in Figure~\ref{fig:question_pie_topic}, the 13 topic categories are: \textit{Medical Complications} (28.6\% of topic annotations), \textit{Reproductive Health} (26.1\%), \textit{Lifestyle \& Daily Living} (20.2\%), \textit{Pre-Transplant} (15.8\%), \textit{Medications} (9.9\%), \textit{Monitoring \& Follow-up} (9.7\%), \textit{Mental \& Emotional Health} (9.6\%), \textit{Surgery \& Recovery} (7.0\%), \textit{Special Populations \& Education} (6.5\%), and four smaller categories covering transplant logistics, financial issues, and community support. Questions are multi-labeled: a single question may be annotated with more than one topic and sub-topic to reflect cross-cutting concerns.

General-type questions account for the largest share of the benchmark (27.9\%; Figure~\ref{fig:question_pie_organ}) and address topics that span organ types, such as immunosuppressant side effects, reproductive health after transplantation, and mental health. At generation time, these questions are answered by \emph{every} handbook in the corpus, while organ-specific questions are answered only by handbooks of the matching organ type. This design ensures comprehensive cross-center coverage for broadly relevant topics while maintaining clinical specificity for organ-level questions.

Questions were lightly paraphrased for anonymization and to ensure they are self-contained (i.e., interpretable without conversational context). The original sources were predominantly U.S.-based (69.9\% geolocated to the United States), consistent with the U.S.\ institutional focus of the handbook corpus. The benchmark was drawn from a larger pool of over 3{,}000 candidate questions collected across organ-specific sources, which were filtered for relevance, deduplication, and quality to produce the final set of 1{,}115.

\subsection{Document Extraction}
\label{sec:extraction}

\begin{figure*}[htbp]
    \centering
    \includegraphics[width=.9\textwidth, trim={1cm 1cm 1cm 1cm}, clip]{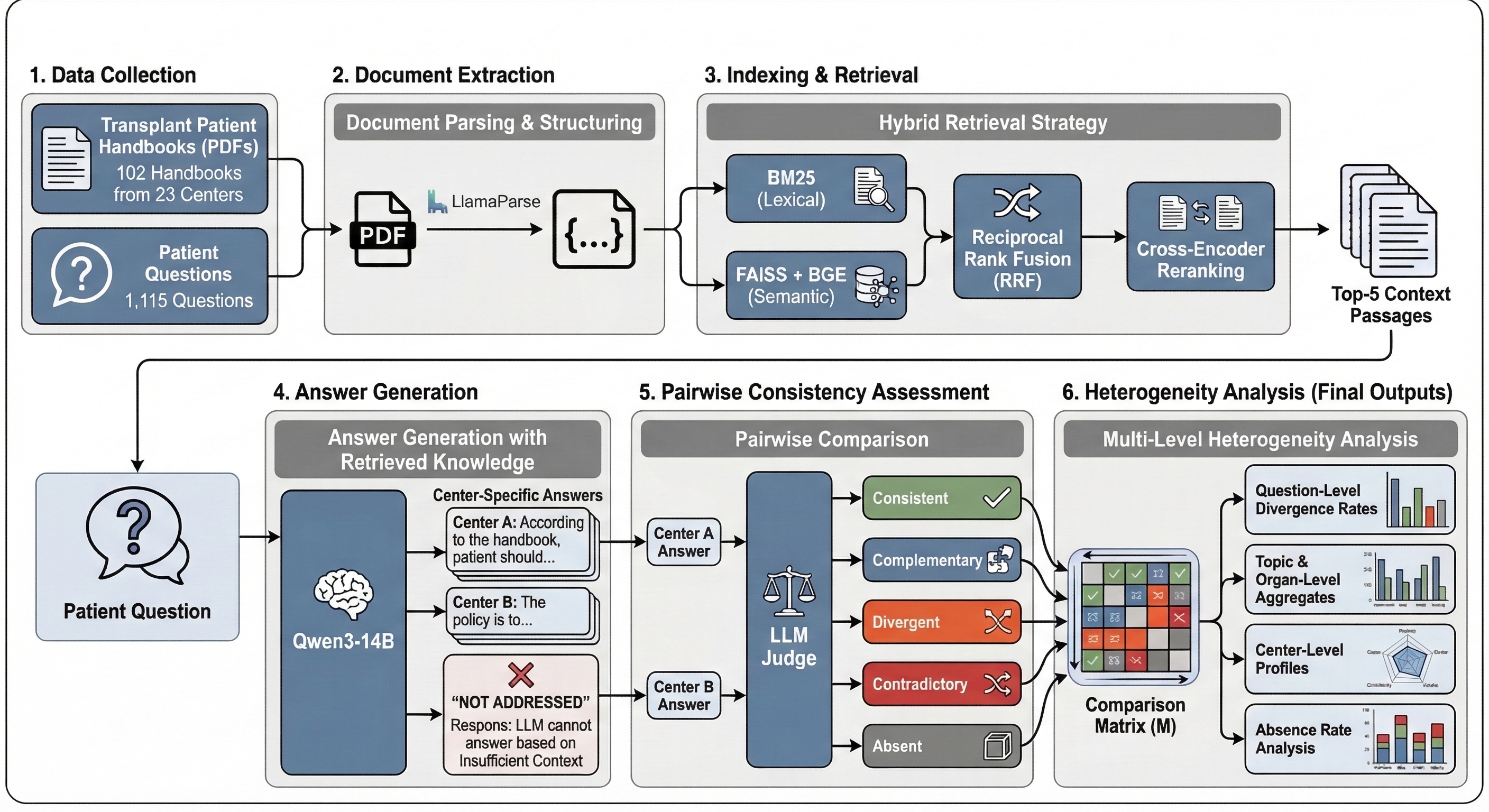}
    \caption{Overview of the experimental pipeline.}
    \label{fig:pipeline}
\end{figure*}

Figure~\ref{fig:pipeline} illustrates the end-to-end pipeline. Raw PDF handbooks are converted to structured JSON representations using LlamaParse~\citep{llamaindex_llamaparse}, a document parsing service that preserves section headings, paragraph boundaries, and page metadata. The output for each handbook is a JSON object containing: organ type, center name, care phase, source file path, full text, and a list of sections---each with its heading, body text, and page numbers. This structured extraction enables section-aware chunking in the downstream retrieval stage (Section~\ref{sec:retrieval}). The extraction pipeline is idempotent: already-processed files are detected and skipped, ensuring resume-safe execution across incremental corpus updates.

\subsection{Retrieval and Answer Generation}
\label{sec:retrieval}

To ground LLM-generated answers in center-specific content, we adopt a hybrid retrieval strategy that combines sparse lexical matching with dense semantic retrieval, followed by neural reranking.

\textit{Indexing.} Leveraging the structured JSON output from the extraction stage (Section~\ref{sec:extraction}), each handbook is segmented into chunks at section boundaries. Sections exceeding 512 tokens are further split at sentence boundaries, with each sub-chunk inheriting the parent section heading as a prefix to preserve topical coherence. All chunks are indexed in two parallel structures: a BM25~\cite{robertson2009probabilistic} inverted index for sparse lexical retrieval and a FAISS~\citep{douze2025faiss} vector index using the \texttt{BAAI/bge-large-en-v1.5} dense encoder~\citep{xiao2024c} for semantic retrieval.

\textit{Hybrid Retrieval with Reranking.} At query time, both retrievers independently return ranked candidate lists. These are merged using Reciprocal Rank Fusion (RRF)~\citep{cormack2009reciprocal} with $k_{\mathrm{RRF}}=60$, which balances the complementary strengths of lexical and semantic matching. The fused candidate set is then reranked with a cross-encoder~\citep{wang2020minilm}), and the top-5 passages are selected as the retrieval context. This hybrid approach has been shown to outperform either retrieval modality alone on biomedical text~\citep{lin2021pyserini}.

\textit{Answer Generation.}
For each question--handbook pair, the top-5 retrieved passages are provided alongside the original question to the generation model (Qwen3-14B at temperature~0), which produces a grounded answer. The model is instructed to rely exclusively on the provided context and to return a standardized \texttt{NOT ADDRESSED} response if the handbook does not contain information relevant to the question, rather than hallucinate an answer.

\subsection{Heterogeneity Analysis}
\label{sec:heterogeneity}

Given the set of center-specific answers generated for each question, we analyze cross-center heterogeneity at multiple levels of granularity. We first identify coverage gaps through absence detection, then classify the relationship between every pair of center responses, and finally aggregate divergence and consistency metrics at the question, topic, organ, and center levels.

\subsubsection{Absence Detection and Coverage Analysis}
\label{sec:absence}

Before comparing center responses, each answer is screened for \emph{absence}---whether the handbook failed to address the question. A fast heuristic first checks for the canonical \texttt{NOT ADDRESSED} prefix produced by the generation model. Answers not matching this prefix are passed to a secondary LLM-based binary classifier (Qwen3-14B) that determines via a structured YES/NO prompt whether the response substantively indicates non-coverage. Absence status is cached per handbook--question pair to avoid redundant inference.

For each organ type and topic category, we report the \emph{absence rate}: the fraction of question--center pairs for which the handbook did not address the question. Let $\mathcal{Q}_g$ denote the set of questions in group $g$ (an organ type or topic category), and let $\mathcal{C}(q)$ denote the set of centers whose handbooks are queried for question $q$. The absence rate for group $g$ is:
\begin{equation}
    r_{\mathrm{abs}}(g) = \frac{\sum_{q \in \mathcal{Q}_g} \sum_{c \in \mathcal{C}(q)} \mathbf{1}[\text{absent}(q, c)]}{\sum_{q \in \mathcal{Q}_g} |\mathcal{C}(q)|}
    \label{eq:abs_rate}
\end{equation}
where $\mathbf{1}[\text{absent}(q, c)]$ is an indicator for whether center $c$'s handbook does not address question $q$. High absence rates indicate systematic coverage gaps---topics that patients ask about but that institutional materials do not address.

\subsubsection{Pairwise Consistency Assessment}
\label{sec:pairwise}

\begin{table}[h]
\centering
\caption{Five-label taxonomy for pairwise comparison of center-specific answers.}
\label{tab:taxonomy}
\renewcommand{\arraystretch}{1.15}
\small
\begin{tabular}{@{}l p{5.5cm} p{6.5cm}@{}}
\toprule
\textbf{Label} & \textbf{Definition} & \textbf{Example} \\
\midrule
\textsc{Absent} & One or both answers indicate the handbook does not address the topic. & Center~A provides dietary guidance; Center~B's handbook contains no relevant section. \\
\textsc{Consistent} & Both answers provide the same clinical recommendation. & Both centers advise avoiding grapefruit due to tacrolimus interactions. \\
\textsc{Complementary} & Clinically compatible but differing in detail or scope. & Center~A lists side effects; Center~B additionally describes management strategies. \\
\textsc{Divergent} & Substantive, clinically meaningful differences (e.g., different thresholds or timelines). & Center~A recommends exercise at 6~weeks post-transplant; Center~B at 8--12~weeks. \\
\textsc{Contradictory} & Directly opposing clinical guidance. & Center~A allows ABO-incompatible live donors; Center~B states they cannot proceed. \\
\bottomrule
\end{tabular}
\end{table}

\textit{Five-Label Comparison Taxonomy.} For each pair of non-absent answers from two different centers, an LLM judge classifies their relationship into one of five categories. Table~\ref{tab:taxonomy} defines each label and provides an illustrative example.

Pairs involving at least one absent answer are assigned \textsc{Absent} without further LLM inference. For all remaining (non-absent) pairs, the LLM judge is prompted with the original question and both center answers, and returns a structured JSON object containing: the classification label, a 2--3 sentence clinical justification, a divergence sub-topic tag (if applicable), and a clinical significance rating (low, medium, or high) for pairs labeled \textsc{Divergent} or \textsc{Contradictory}. Inference is performed with greedy decoding (temperature~0) for reproducibility.

\textit{Comparison Matrix.} For a question answered by $N$ centers, the $\binom{N}{2}$ unique pairwise comparisons form a symmetric $N \times N$ comparison matrix $\mathbf{M}$, where entry $M_{ij}$ records the label assigned to the pair of handbooks $i$ and $j$. Diagonal entries are defined as \textsc{Consistent} by convention. Each pairwise result is persisted independently as a JSON file, enabling resume-safe incremental execution over the full corpus.

\subsubsection{Question-Level Heterogeneity}
\label{sec:question_level}

For each question $q$, we quantify cross-center agreement and disagreement using two complementary metrics computed over all $\binom{N}{2}$ center pairs.

The \emph{divergence rate} measures the fraction of non-absent pairs exhibiting clinically meaningful disagreement:
\begin{equation}
    r_{\mathrm{div}}(q) = \frac{|\{(i,j) : M_{ij}^{(q)} \in \{\textsc{Divergent}, \textsc{Contradictory}\}\}|}{|\{(i,j) : M_{ij}^{(q)} \neq \textsc{Absent}\}|}
    \label{eq:div_rate}
\end{equation}

The \emph{consistency rate} measures the fraction of non-absent pairs in full agreement:
\begin{equation}
    r_{\mathrm{con}}(q) = \frac{|\{(i,j) : M_{ij}^{(q)} = \textsc{Consistent}\}|}{|\{(i,j) : M_{ij}^{(q)} \neq \textsc{Absent}\}|}
    \label{eq:con_rate}
\end{equation}

where $M_{ij}^{(q)}$ is the comparison label for question $q$ between centers $i$ and $j$. Questions with high $r_{\mathrm{div}}$ identify topics where institutional guidance is most fragmented, while questions with high $r_{\mathrm{con}}$ indicate consensus across responding centers. Note that $r_{\mathrm{div}}(q) + r_{\mathrm{con}}(q) \leq 1$, with the residual fraction accounted for by \textsc{Complementary} pairs.

\subsubsection{Topic- and Organ-Level Aggregation}
\label{sec:topic_organ}

Both metrics are aggregated by topic category and organ type. For a group $g$ (topic or organ) with associated question set $\mathcal{Q}_g$, the group-level divergence and consistency rates are:
\begin{equation}
    R_{\mathrm{div}}(g) = \frac{1}{|\mathcal{Q}_g|} \sum_{q \in \mathcal{Q}_g} r_{\mathrm{div}}(q), \qquad
    R_{\mathrm{con}}(g) = \frac{1}{|\mathcal{Q}_g|} \sum_{q \in \mathcal{Q}_g} r_{\mathrm{con}}(q)
    \label{eq:group_rates}
\end{equation}

We additionally report the proportion of questions within each group for which $r_{\mathrm{div}} > 0$ (i.e., at least one divergent or contradictory pair exists). This two-metric approach distinguishes between groups where divergence is pervasive (many questions affected) and groups where it is concentrated in a few high-disagreement questions.

\subsubsection{Center-Level Profiles}
\label{sec:center_level}

For each center $c$, we compute a heterogeneity profile by aggregating pairwise labels across all questions and partner centers. Let $\mathcal{P}(c)$ denote the set of all non-absent pairwise comparisons involving center $c$. The center-level divergence and consistency rates are:
\begin{equation}
    R_{\mathrm{div}}(c) = \frac{|\{(q,j) \in \mathcal{P}(c) : M_{cj}^{(q)} \in \{\textsc{Divergent}, \textsc{Contradictory}\}\}|}{|\mathcal{P}(c)|}
    \label{eq:center_div}
\end{equation}
\begin{equation}
    R_{\mathrm{con}}(c) = \frac{|\{(q,j) \in \mathcal{P}(c) : M_{cj}^{(q)} = \textsc{Consistent}\}|}{|\mathcal{P}(c)|}
    \label{eq:center_con}
\end{equation}

Centers with consistently high $R_{\mathrm{div}}$ across topics may reflect systematically different institutional policies, while centers with high $R_{\mathrm{con}}$ relative to peers indicate alignment with prevailing practice norms. These profiles enable identification of outlier institutions whose guidance departs most from the cross-center consensus.

\section{Results}

We applied the full pipeline to the corpus of 102 handbooks from 23 centers across five organ types, using all 1{,}115 benchmark questions. This section reports heterogeneity findings organized by global label distribution, coverage gaps, organ- and topic-level divergence, center-level profiles, and illustrative matrix visualizations.

\subsection{Global Label Distribution}

Across 1{,}115 questions and 102 handbooks, the pairwise comparison pipeline produced 1{,}772{,}261 handbook pairs. Of these, 1{,}704{,}242 (96.2\%) were classified as \textsc{Absent}, reflecting the expected sparsity: most handbooks address only a subset of the benchmark questions. Among the 68{,}019 non-absent pairs (i.e., pairs where both handbooks substantively addressed the question), the label distribution was: \textsc{Complementary} 44{,}870 (66.0\%), \textsc{Divergent} 14{,}132 (20.8\%), \textsc{Consistent} 8{,}874 (13.0\%), and \textsc{Contradictory} 143 (0.2\%). Table~\ref{tab:global_labels} summarizes the global distribution.

\begin{table}[H]
\centering
\caption{Global pairwise label distribution across all 1{,}115 questions. Percentages in the right column are computed over non-absent pairs only.}
\label{tab:global_labels}
\small
\begin{tabular}{lrr}
\toprule
\textbf{Label} & \textbf{Count} & \textbf{\% of non-absent} \\
\midrule
\textsc{Absent}        & 1{,}704{,}242 & --- \\
\textsc{Complementary} & 44{,}870      & 66.0\% \\
\textsc{Divergent}     & 14{,}132      & 20.8\% \\
\textsc{Consistent}    & 8{,}874       & 13.0\% \\
\textsc{Contradictory} & 143           & 0.2\% \\
\midrule
\textbf{Total non-absent} & \textbf{68{,}019} & \textbf{100\%} \\
\bottomrule
\end{tabular}
\end{table}

The dominance of \textsc{Complementary} labels indicates that when two centers both address a question, they most frequently provide compatible but differently scoped information. However, one in five non-absent pairs exhibits clinically meaningful divergence, and a small but non-negligible number involve direct contradictions.

\subsection{Coverage Gaps: Absence Rates by Organ and Topic}
\label{sec:results_absence}

Absence rates varied substantially across organ types and topic categories (Table~\ref{tab:absence}). General-type questions, which are posed to all 102 handbooks, had the highest absence rate (90.5\%), reflecting the fact that organ-specific handbooks rarely cover cross-cutting topics such as reproductive health or mental wellness. Among organ-specific question sets, lung had the lowest absence rate (72.5\%), suggesting broader topical coverage in lung transplant handbooks, while pancreas had the highest (85.6\%).

\begin{table}[h]
\centering
\caption{Absence rates by organ type and selected topic categories. The absence rate is the fraction of question--handbook pairs in which the handbook did not address the question (Eq.~\ref{eq:abs_rate}).}
\label{tab:absence}
\small
\begin{tabular}{lrrr}
\toprule
\textbf{Group} & \textbf{Questions} & \textbf{Pairs} & \textbf{Absence rate} \\
\midrule
\multicolumn{4}{l}{\textit{By organ type}} \\
General   & 311 & 31{,}722 & 0.905 \\
Heart     & 137 &  3{,}562 & 0.818 \\
Kidney    & 196 &  4{,}312 & 0.780 \\
Liver     & 164 &  2{,}788 & 0.805 \\
Lung      & 153 &  3{,}978 & 0.725 \\
Pancreas  & 154 &  1{,}694 & 0.856 \\
\midrule
\multicolumn{4}{l}{\textit{By topic (selected)}} \\
Reproductive Health              & 291 & --- & 0.951 \\
Transplant Process \& Logistics  &  21 & --- & 0.894 \\
Medications                      & 110 & --- & 0.838 \\
Mental \& Emotional Health       & 107 & --- & 0.829 \\
Monitoring \& Follow-up          & 108 & --- & 0.785 \\
Financial \& Administrative      &  19 & --- & 0.724 \\
\bottomrule
\end{tabular}
\end{table}

Reproductive Health had the highest topic-level absence rate (95.1\%), indicating an important gap in patient education materials: the vast majority of handbooks do not address fertility, contraception, or pregnancy after transplant. Financial \& Administrative topics had the lowest absence rate (72.4\%).

\subsection{Organ-Level Heterogeneity}
\label{sec:results_organ}

Table~\ref{tab:organ_metrics} reports organ-level divergence and consistency rates (Eq.~\ref{eq:group_rates}). Kidney and lung exhibited the highest divergence: 39.8\% and 41.8\% of questions had at least one divergent or contradictory pair, respectively. Pancreas showed the lowest divergence prevalence (11.0\%), likely reflecting both fewer centers and more standardized guidance for pancreas transplantation.

\begin{table}[H]
\centering
\caption{Organ-level heterogeneity metrics. $Q_{\text{total}}$: total questions for the organ; $Q_{\text{active}}$: questions with $\geq$1 non-absent pair; $R_{\text{div}}$ and $R_{\text{con}}$: mean question-level divergence and consistency rates (Eq.~\ref{eq:group_rates}); \%Div: percentage of questions with at least one divergent or contradictory pair.}
\label{tab:organ_metrics}
\small
\begin{tabular}{lrrrrrr}
\toprule
\textbf{Organ} & $Q_{\text{total}}$ & $Q_{\text{active}}$ & $R_{\text{div}}$ & $R_{\text{con}}$ & \textbf{\%Div} \\
\midrule
General   & 311 & 122 & 0.188 & 0.285 & 28.6\% \\
Heart     & 137 &  69 & 0.152 & 0.186 & 26.3\% \\
Kidney    & 196 & 124 & 0.237 & 0.148 & 39.8\% \\
Liver     & 164 &  91 & 0.209 & 0.142 & 30.5\% \\
Lung      & 153 & 117 & 0.152 & 0.136 & 41.8\% \\
Pancreas  & 154 &  44 & 0.195 & 0.223 & 11.0\% \\
\bottomrule
\end{tabular}
\end{table}

General-type questions had the highest mean consistency rate ($R_{\text{con}} = 0.285$), suggesting that cross-cutting topics such as immunosuppression adherence and lifestyle guidance tend to elicit more uniform recommendations when they are covered. Lung and liver had the lowest consistency rates (0.136 and 0.142), indicating that even when centers address the same organ-specific question, they often differ in the specifics of their recommendations.

\subsection{Topic-Level Heterogeneity}
\label{sec:results_topic}

Table~\ref{tab:topic_metrics} presents heterogeneity metrics for the 13 topic categories. Monitoring \& Follow-up exhibited the highest mean divergence rate ($R_{\text{div}} = 0.277$) with 38.9\% of questions showing at least one divergent pair, reflecting well-documented variation in post-transplant surveillance protocols across centers. Lifestyle \& Daily Living followed ($R_{\text{div}} = 0.235$, 40.4\% divergence prevalence), consistent with the lack of standardized guidelines for diet, exercise, and activity restrictions.

\begin{table}[h]
\centering
\caption{Topic-level heterogeneity metrics (topics with $\geq$20 questions). Columns as in Table~\ref{tab:organ_metrics}.}
\label{tab:topic_metrics}
\small
\begin{tabular}{lrrrrrr}
\toprule
\textbf{Topic} & $Q_{\text{total}}$ & $Q_{\text{active}}$ & $R_{\text{div}}$ & $R_{\text{con}}$ & \textbf{\%Div} \\
\midrule
Monitoring \& Follow-up        & 108 &  59 & 0.277 & 0.073 & 38.9\% \\
Lifestyle \& Daily Living      & 225 & 134 & 0.235 & 0.128 & 40.4\% \\
Pre-Transplant                 & 176 & 105 & 0.221 & 0.179 & 29.5\% \\
Medical Complications          & 319 & 168 & 0.179 & 0.177 & 26.6\% \\
Reproductive Health            & 291 &  85 & 0.176 & 0.315 & 17.5\% \\
Special Populations            &  73 &  39 & 0.172 & 0.242 & 24.7\% \\
Surgery \& Recovery            &  78 &  36 & 0.154 & 0.098 & 29.5\% \\
Financial \& Administrative    &  19 &  10 & 0.147 & 0.216 & 36.8\% \\
Medications                    & 110 &  63 & 0.120 & 0.233 & 30.0\% \\
Mental \& Emotional Health     & 107 &  51 & 0.087 & 0.193 & 19.6\% \\
\bottomrule
\end{tabular}
\end{table}

Conversely, Reproductive Health showed the highest consistency rate ($R_{\text{con}} = 0.315$) but the lowest divergence prevalence (17.5\%). This initially surprising finding reflects its extreme absence rate (95.1\%): the few centers that do address reproductive topics tend to convey similar core messages (e.g., avoid pregnancy in the first year), but the vast majority of handbooks omit this information entirely. Mental \& Emotional Health had the lowest divergence ($R_{\text{div}} = 0.087$), suggesting that psychosocial guidance, when provided, is relatively uniform.

\subsection{Center-Level Heterogeneity}
\label{sec:results_center}

Among the 23 anonymized centers with sufficient data for analysis, center-level divergence rates ranged from $R_{\text{div}} = 0.139$ to $0.255$, and consistency rates from $R_{\text{con}} = 0.082$ to $0.194$. The three centers with the highest divergence profiles (center-024, center-019, center-008; $R_{\text{div}} \geq 0.238$) consistently produced answers differing from peer institutions, while the most consistent centers (center-022, center-012; $R_{\text{con}} \geq 0.183$) showed stronger alignment with cross-center norms. This variation across centers was not simply an artifact of sample size: the number of non-absent pairs per center ranged from 764 to 12{,}323, and divergence patterns persisted after controlling for data volume.

\subsection{Illustrative Comparison Matrices}
\label{sec:results_matrices}

\begin{figure*}[htbp]
    \centering
    \begin{subfigure}[t]{0.32\textwidth}
        \centering
        \includegraphics[width=\textwidth]{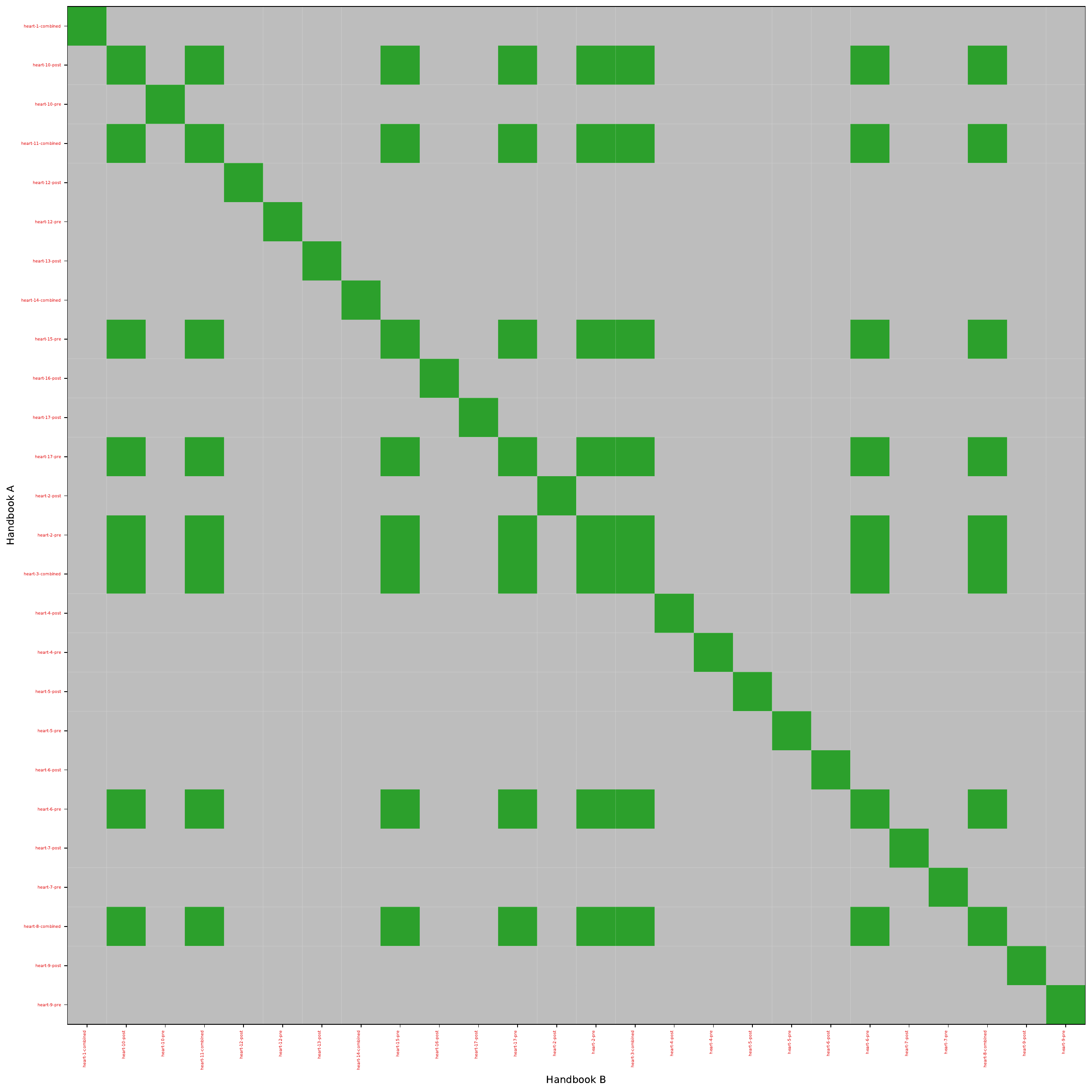}
        \caption{Q22 (heart): ``Does blood type affect how long you wait for a heart transplant?'' All 28 non-absent pairs are \textsc{Consistent}---a clear cross-center consensus.}
        \label{fig:hm_q22}
    \end{subfigure}
    \hfill
    \begin{subfigure}[t]{0.32\textwidth}
        \centering
        \includegraphics[width=\textwidth]{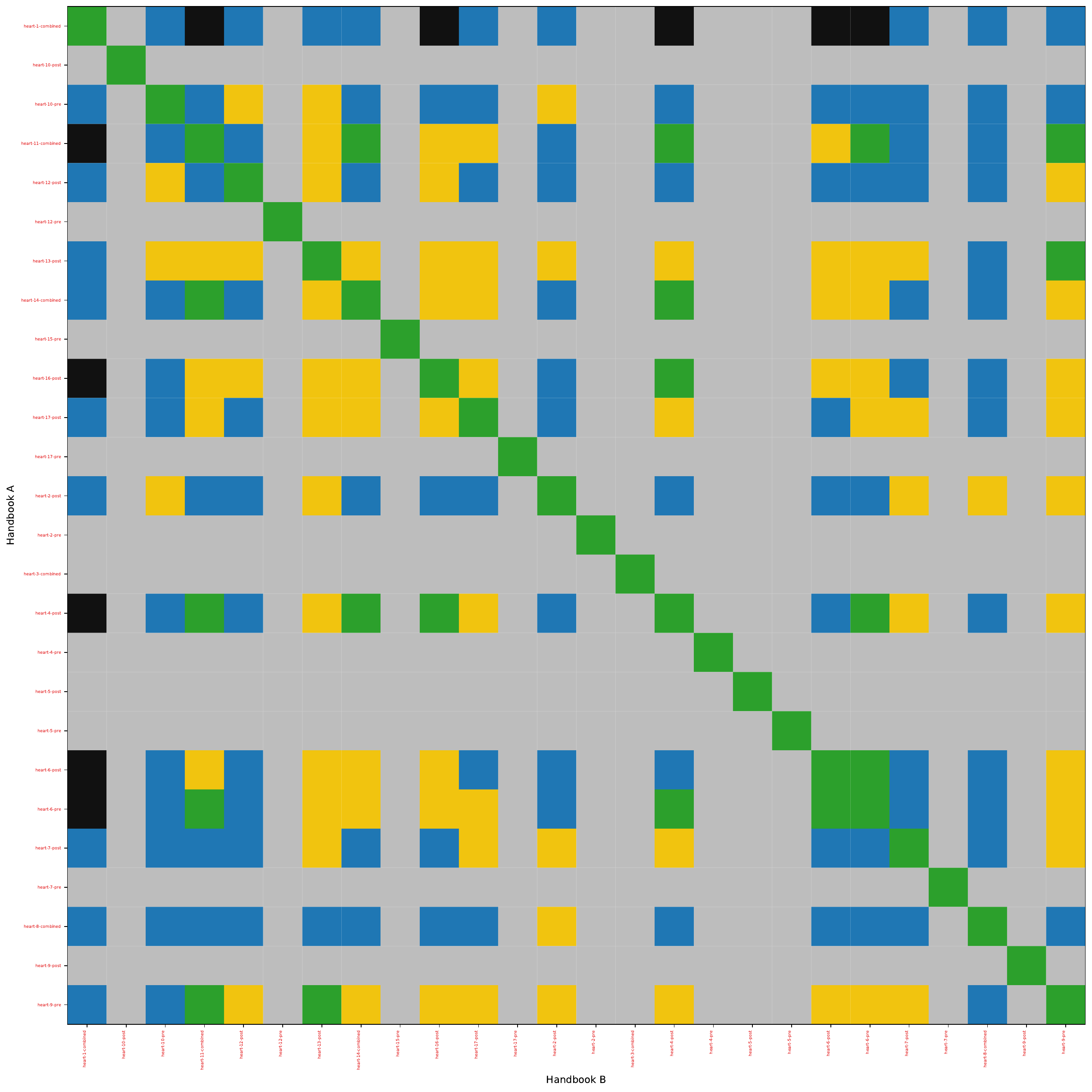}
        \caption{Q219 (heart): ``How should I handle dental care after my transplant?'' All five labels present including \textsc{Contradictory} (black), reflecting known clinical controversy over antibiotic prophylaxis.}
        \label{fig:hm_q219}
    \end{subfigure}
    \hfill
    \begin{subfigure}[t]{0.32\textwidth}
        \centering
        \includegraphics[width=\textwidth]{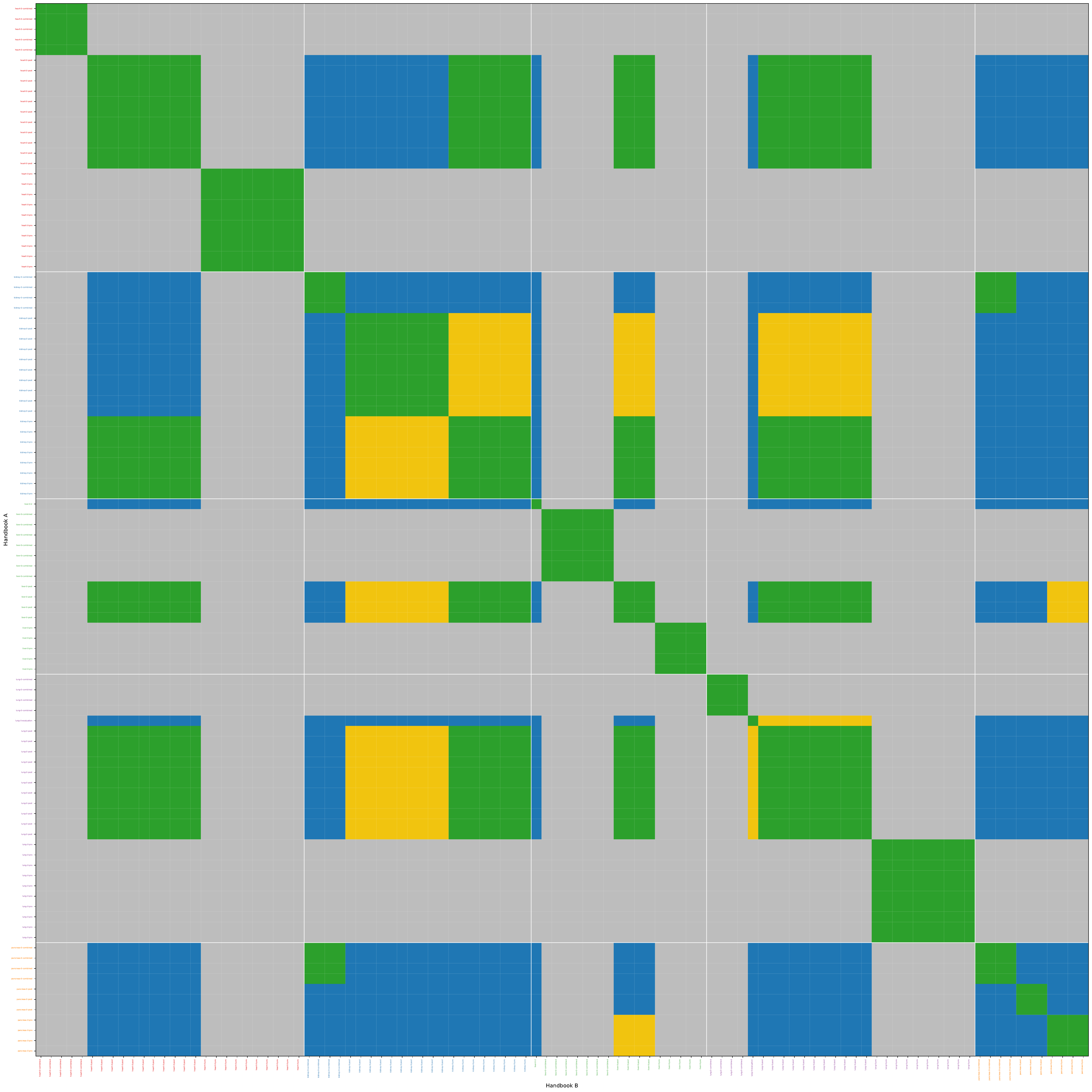}
        \caption{Q46 (general): ``How often should I have blood work done?'' $r_{\text{div}} = 0.871$: centers recommend vastly different monitoring schedules.}
        \label{fig:hm_q46}
    \end{subfigure}

    \caption{Selected comparison matrices illustrating heterogeneity patterns. Each cell represents a pairwise comparison between two center handbooks for a single question. Colors: grey = \textsc{Absent}, green = \textsc{Consistent}, yellow = \textsc{Complementary}, blue = \textsc{Divergent}, black = \textsc{Contradictory}. Handbook labels on axes are anonymized (organ-center\_index-phase). Matrices are symmetric; diagonal entries are \textsc{Consistent} by convention.}
    \label{fig:heatmaps}
\end{figure*}

Figure~\ref{fig:heatmaps} presents three comparison matrices selected to illustrate the range of heterogeneity patterns observed across questions.

These matrices reveal several structural patterns. First, consistency tends to cluster within organ types: in Q22 (Figure~\ref{fig:hm_q22}), all responding heart centers agree on the role of blood type in waitlist priority, reflecting well-established allocation policy. Second, divergence can emerge even in ostensibly straightforward questions: Q46 (Figure~\ref{fig:hm_q46}) asks about blood work frequency, yet centers recommend schedules ranging from weekly to every few months, producing $r_{\text{div}} = 0.871$---the highest in the dataset. Third, the coexistence of all five labels within a single matrix (Figure~\ref{fig:hm_q219}) demonstrates that institutional disagreement can be highly structured: for dental care after heart transplant, specific center pairs consistently contradict one another on antibiotic prophylaxis while agreeing with other peers. More broadly, \textsc{Complementary} patterns---in which centers provide medically compatible guidance that differs in scope and emphasis---are the most common non-absent label across the dataset, reflecting institutional preferences in what to highlight for patients.

\section{Discussion and Conclusion}

\subsection{Principal Findings}

% This study introduces a systematic, scalable framework for measuring institutional heterogeneity in transplant patient education materials using retrieval-augmented language models. Our analysis of 102 handbooks from 23 U.S.\ transplant centers reveals several key findings.

First, \textbf{cross-center divergence is significant but unevenly distributed.} Among 68{,}019 non-absent pairwise comparisons, 20.8\% were classified as \textsc{Divergent} and 0.2\% as \textsc{Contradictory}. Lung and kidney questions showed the highest divergence prevalence (41.8\% and 39.8\% of questions), while pancreas showed the lowest (11.0\%). At the topic level, Monitoring \& Follow-up and Lifestyle \& Daily Living exhibited the highest divergence rates ($R_{\text{div}} = 0.277$ and $0.235$), consistent with the absence of standardized national guidelines for post-transplant surveillance and activity restrictions.

Second, \textbf{coverage gaps are a dominant source of information inequality.} The overall absence rate of 96.2\% indicates that most handbooks address only a narrow slice of patients' information needs. Reproductive Health had a 95.1\% absence rate, meaning the vast majority of handbooks provide no guidance on fertility, contraception, or pregnancy. This systematic omission may be more consequential than outright disagreement, as patients at certain centers receive no information on an important topics rather than merely different information.

Third, \textbf{center-level divergence profiles are stable and interpretable.} Divergence rates ranged from 0.139 to 0.255 across centers, suggesting that some institutions systematically depart from cross-center norms in ways that reflect genuine differences in clinical philosophy or authoring conventions rather than sample size artifacts.

Fourth, \textbf{individual questions exhibit structured heterogeneity.} The comparison matrices (Figure~\ref{fig:heatmaps}) show that divergence is not random noise: topics such as antibiotic prophylaxis for dental care (Q219) and monitoring frequency (Q46) produce structured patterns of agreement and disagreement reflecting documented clinical controversies. These patterns underscore that document selection in RAG-based medical question answering functions as an implicit clinical decision, as systems grounded in a single center's materials inherit that institution's omissions and positions on contested topics. The coverage gap analysis and center-level profiles can directly support quality improvement by identifying topics where education materials should be expanded and flagging institutions most misaligned with peer consensus. More broadly, the framework is applicable to other domains where multiple institutions issue guidance on overlapping topics, including oncology protocols, chronic disease management, and discharge instructions.

\subsection{Limitations}

Several limitations should be acknowledged. Although we conducted sample annotation and agreement checks, the LLM-based pairwise judge may still introduce systematic classification biases. The hybrid retrieval pipeline, while effective, is not perfect; more advanced retrieval methods could improve passage selection and reduce noise in the generated answers. Additionally, our pipeline currently processes only the textual content of handbooks, yet many handbooks also contain tables, figures, and infographics that may address questions currently identified as coverage gaps; incorporating multimodal extraction could therefore refine gap estimates. Finally, our corpus is limited to English-language handbooks from 23 U.S.\ transplant centers, and extending the framework to non-English materials and international transplant systems would improve generalizability.

\subsection{Conclusion}

This work introduces a scalable framework for quantifying institutional heterogeneity in transplant patient education materials through document-grounded language model comparison. Our analysis of 102 handbooks from 23 U.S.\ centers reveals that over 20\% of non-absent pairwise comparisons reflect clinically meaningful divergence, concentrated in topics like post-transplant monitoring and lifestyle restrictions, while coverage gaps are even more pervasive, with reproductive health exhibiting a 95.1\% absence rate. These findings highlight that document selection in retrieval-augmented medical question-answering task functions as an implicit clinical decision, reinforcing current practice. Systems grounded in a single center's materials inherit that institution's omissions and positions on difficult and contested topics. The topic-level absence and center-level divergence profiles offer actionable benchmarks for harmonizing patient education. By making cross-center variation measurable and structured, our framework provides a foundation for more transparent and comprehensive  document-grounded medical question answering to meet the diverse information needs of patients with complex health conditions.

\section*{Acknowledgments}
This work used Bridges-2 at the Pittsburgh Supercomputing Center (PSC) through allocation CIS250181 from the \href{https://access-ci.org/}{Advanced Cyberinfrastructure Coordination Ecosystem: Services \& Support} (ACCESS) program, which is supported by U.S. National Science Foundation grants \#2138259, \#2138286, \#2138307, \#2137603, and \#2138296. We sincerely thank all the transplant centers that provided their patient handbooks for this analysis. We are grateful to T. Mace and J. Mace at Transplants.org who assembled and shared the handbooks with us for this study. This research was also funded by the National Institute of Standards and Technology under Federal Award ID Number 60NANB24D231 and Carnegie Mellon University AI Measurement Science and Engineering Center (AIMSEC).

%\references
% References as numbers
\renewcommand{\bibsection}{\centering\section*{\refname}}
\makeatletter
\renewcommand{\@biblabel}[1]{\hfill #1.}
\makeatother

% unstr is used to keep citation order
\bibliographystyle{vancouver}
\bibliography{amia}  

@misc{llamaindex_llamaparse,
  title        = {LlamaParse},
  author       = {LlamaIndex},
  year         = {2024},
  howpublished = {\url{https://developers.llamaindex.ai/}},
  note         = {Document parsing platform for LLM applications}
}

@book{robertson2009probabilistic,
  title={The probabilistic relevance framework: BM25 and beyond},
  author={Robertson, Stephen and Zaragoza, Hugo},
  volume={4},
  year={2009},
  publisher={Now Publishers Inc}
}

@article{douze2025faiss,
  title={The faiss library},
  author={Douze, Matthijs and Guzhva, Alexandr and Deng, Chengqi and Johnson, Jeff and Szilvasy, Gergely and Mazar{\'e}, Pierre-Emmanuel and Lomeli, Maria and Hosseini, Lucas and J{\'e}gou, Herv{\'e}},
  journal={IEEE Transactions on Big Data},
  year={2025},
  publisher={IEEE}
}

@inproceedings{xiao2024c,
  title={C-pack: Packed resources for general chinese embeddings},
  author={Xiao, Shitao and Liu, Zheng and Zhang, Peitian and Muennighoff, Niklas and Lian, Defu and Nie, Jian-Yun},
  booktitle={Proceedings of the 47th international ACM SIGIR conference on research and development in information retrieval},
  pages={641--649},
  year={2024}
}

@inproceedings{cormack2009reciprocal,
  title={Reciprocal rank fusion outperforms condorcet and individual rank learning methods},
  author={Cormack, Gordon V and Clarke, Charles LA and Buettcher, Stefan},
  booktitle={Proceedings of the 32nd international ACM SIGIR conference on Research and development in information retrieval},
  pages={758--759},
  year={2009}
}

@article{wang2020minilm,
  title={Minilm: Deep self-attention distillation for task-agnostic compression of pre-trained transformers},
  author={Wang, Wenhui and Wei, Furu and Dong, Li and Bao, Hangbo and Yang, Nan and Zhou, Ming},
  journal={Advances in neural information processing systems},
  volume={33},
  pages={5776--5788},
  year={2020}
}

@article{rivera2025examining,
  title={Examining Transparency in Kidney Transplant Recipient Selection Criteria: Nationwide Cross-Sectional Study},
  author={Rivera, Belen and Canizares, Stalin and Cojuc-Konigsberg, Gabriel and Holub, Olena and Nakonechnyi, Alex and Chumdermpadetsuk, Ritah R and Ladin, Keren and Eckhoff, Devin E and Allen, Rebecca and Pawar, Aditya},
  journal={JMIR AI},
  volume={4},
  pages={e74066},
  year={2025},
  publisher={JMIR Publications Toronto, Canada}
}

@article{king2020major,
  title={Major variation across local transplant centers in probability of kidney transplant for wait-listed patients},
  author={King, Kristen L and Husain, S Ali and Schold, Jesse D and Patzer, Rachel E and Reese, Peter P and Jin, Zhezhen and Ratner, Lloyd E and Cohen, David J and Pastan, Stephen O and Mohan, Sumit},
  journal={Journal of the American Society of Nephrology},
  volume={31},
  number={12},
  pages={2900--2911},
  year={2020},
  publisher={LWW}
}

@article{mace2025improving,
  title={Improving Quality of Patient Educational Materials through a Comparative Analysis of Patient Handbooks from US Transplant Centers},
  author={Mace, T and Mace, J and Friedman, B and Padman, R and Clemente, S and Abdulakhadov, A and An, J and Duken, J and John, S and Kentilitisca, Y and others},
  journal={American Journal of Transplantation},
  volume={25},
  number={8},
  pages={S995},
  year={2025},
  publisher={Elsevier}
}

@article{rodrigue2017readability,
  title={Readability, content analysis, and racial/ethnic diversity of online living kidney donation information},
  author={Rodrigue, James R and Feranil, Mario and Lang, Jenna and Fleishman, Aaron},
  journal={Clinical transplantation},
  volume={31},
  number={9},
  pages={e13039},
  year={2017},
  publisher={Wiley Online Library}
}

@article{poudel2024readability,
  title={Readability of online patient education materials related to liver transplantation in the United States},
  author={Poudel, Ayusha and Adhikari, Anurag and Poudel, Sajana and Poudel, Aayush},
  journal={Transplantology},
  volume={5},
  number={3},
  pages={216--223},
  year={2024},
  publisher={MDPI}
}

@inproceedings{lin2021pyserini,
  title={Pyserini: A Python Toolkit for Reproducible Information Retrieval Research with Sparse and Dense Representations},
  author={Lin, Jimmy and Ma, Xueguang and Lin, Sheng-Chieh and Yang, Jheng-Hong and Pradeep, Ronak and Nogueira, Rodrigo},
  booktitle={Proceedings of the 44th International ACM SIGIR Conference on Research and Development in Information Retrieval (SIGIR)},
  year={2021}
}

\end{document}